\documentclass[a4paper, 10pt]{article}    
\usepackage{latexsym}
\usepackage{amssymb}
\usepackage{amsmath}
\usepackage{amsfonts}
\usepackage{graphicx}
\usepackage{float}
\usepackage[english]{babel}
\usepackage{array} 

\usepackage{changebar}
\usepackage{xcolor}

\title{ \bf Bilateral Credit Valuation Adjustment\\ of an Optional Early Termination Clause}  
\author{ Lorenzo Giada \\ Banco Popolare, Verona \\ {\tt lorenzo.giada@gmail.com\,}$^*$ \and Claudio Nordio \\ Banco Popolare, Verona \\ {\tt c.nordio@gmail.com} \footnote{This paper reflects the authors' opinions and
not necessarily those of their employers.}}

\begin{document}           
\maketitle                 

\begin{abstract}
\it \noindent
Is an option to early terminate a swap at its market value worth zero? At first sight it seems so, but in presence of counterparty risk it depends on the criteria used to determine such market value. In case of a single uncollateralised swap transaction between two defaultable counterparties, the additional unilateral option to early terminate the swap at predefined dates requires the evaluation of a Bermudan option. We give a general pricing formula assuming a default-free close-out amount, and apply it to different kinds of derivatives in a setting with deterministic default intensity, showing that the impact on the fair value at inception and on the counterparty risk of the transaction might be non negligible.
\end{abstract}

\bigskip

{\bf JEL} Classification codes: G12, G13 

{\bf AMS} Classification codes: 62H20, 62L15, 91B25, 91B70 

\bigskip


\vspace{15mm}

\section{Summary}

The impact of close-out conventions in the ISDA setting has been highlighted in \cite{brigo3}, \cite{brigo2} with particular regard to the consequences on bilateral counterparty risk adjustments. The authors show how the contractual definition of the close out amount (``risk-free'' vs ``substitution'') in case of default has a strong impact on the fair price of the transaction at inception; moreover, the ISDA Close-out Amount Protocol extends such conventions to any Additional Early Termination event (ATE), see 6(e)(ii)(1) and pag 17 (par 5) in \cite{isda}. A wide debate \cite{riskmagazine1}, \cite{riskmagazine2} has taken place on the implication of the close-out definition and has focused in particular on the rating-triggered ATEs \cite{mercurio}.

Here, we focus instead on how to value a single uncollateralized transaction where a party has the option to early terminate the transaction at predefined dates by exchanging the default risk-free amount with the other party. This kind of ATE is commonly introduced to shorten the horizon of the credit line between the parties; in the following we will refer to it as a {\it break clause} (BC). This case is different, and to a certain extent simpler, from those examined in previous literature, where the exercise date of the early termination is a priori unknown. We will obtain a general but remarkably simple pricing formula for the case of default-free close out amount, a choice particularly interesting, as it is the only counterparty-independent one, and therefore a good candidate to a less arbitrary close-out convention \cite{riskmagazine4}, even if it is subject to criticism \cite{brigo3} when used to determine the settlement amount in case of default of the counterparty. 

It is clear that a Bermudan-like optionality is in general related to a BC, and that the strike of the option is defined by the close-out amount. Our main result shows how the choice of a default risk-free close-out amount implies a remarkably simple form for the payoff of such an option. On the other hand, the optionality is worthless only if we  assume the close-out amount to be equal to the current value of the transaction, a quite improbable scenario - at least under the ISDA protocol when the close-out amount is determined on the basis of the quotations of third parties whose creditworthiness may be far from that of the original counterparty. Moreover, when the termination event is triggered by a BC, we remark that the choice of a default risk-free amount naturally arises from the ISDA protocol. In fact, on one hand, if the parties have specified that mid-market quotations should be used to determine the termination payment, the default risk of the exercising party does not have to be considered in the quotation (pag 17, par 5 in \cite{isda}); on the other hand, a rational choice for the exercising party that minimizes the CVA that negatively affects the calculation of the termination amount, is to collect quotations from counterparties with high credit standing or no default risk (e.g. clearing members of a central counterparty)\footnote{Such quotations are used to calculate the termination amount, but it is not required under ISDA to replace the transaction with any of the counterparties providing them.}.

First, we will apply the pricing formula to show how it works in the simple case of an equity forward with a unique BC. In this case quasi-analytical results are available, at least assuming deterministic intensities for the default processes, using the well known Geske-Johnson approach for bermudan derivatives (see e.g.~\cite{brockhaus}), which is an effective technique when the exercise dates are few (as is often the case, e.g.~for medium dated swaps, i.e.~with maturity between 7 and 10 years). We will then address a more general case by using the pricing formula on a trinomial tree to evaluate a receiver/payer interest swap with multiple BCs. In all cases we find considerable impacts on the par strike. Finally, we will briefly discuss how a BC may be effective in mitigating the counterparty risk related to a transaction, at least from a Basel III perspective, to the extent that it reduces the variability of the expected loss (the so-called unilateral credit valuation adjustment - UCVA) due to the deterioration of the credit worthiness of the counterparty. Furthermore, a mutual BC is even more effective to mitigate the unilateral CVA risk, since it cancels the contribution to the unilateral CVA after the first exercise date. These results give a strong quantitative support to the capital relief discussion in \cite{riskmagazine5}.

Even if our approach is at a transaction level and cannot easily include the netting effects with other transactions referenced to the same ISDA agreement, we point out that our work has practical implications, due to the wide presence in the market of uncollateralized swap referenced to a single-transaction agreement, or of portfolios composed by a few large transactions with the same side (e.g.~a portfolio of payer interest rate swaps). For instance, consider the case of a derivative between a bank and a corporate, between a bank and a sovereign (see \cite{riskmagazine3} for a  notable example), and of a so-called back-swap linked to securitization or covered bond. These cases often concern long-dated swaps and typically involve BCs.

\section{Notation}
In the following we refer to the notation and the results of \cite{brigo2}. The value $V^{AB}_B(t_0)$ of a derivative contract between two defaultable counterparty $A$ and $B$ as seen from $B$ in $t_0$, when the first to default effect is taken into account, is 
\begin{equation}\label{fairvalue}
V^{AB}_B(t_0) = V^0_B(t_0) - \mathbb{E}\left\{\left. L_A\mathbb{I}_A(t_0,T) D(t_0,\tau_A) \left[ V_B^0(\tau_A) \right]^+
\right| t_0 \right\}
+ \mathbb{E}\left\{\left. L_B\mathbb{I}_B(t_0,T) D(t_0,\tau_B) \left[ -V_B^0(\tau_B) \right]^+ \right| t_0 \right\},
\end{equation}
where $D(t_1,t_2)$ is the stochastic discount factor between two dates, $\tau_X$ is the default time of counterparty $X$, $T$ is the last payment date of the derivative, $L_X$ the loss given default of counterparty $X$, $\mathbb{I}_A(t_1,t_2)=\mathbb{I}_{t_1<\tau_A<\min(\tau_B,t_2)}$, $\mathbb{I}_B(t_1,t_2)=\mathbb{I}_{t_1<\tau_B<\min(\tau_A,t_2)}$ (the last two terms take into account the order of default of the counterparties). $V^0_B(t)$ is the value of the equivalent default-free derivative as seen from $B$,
\begin{equation}
V_B^0(t) = \mathbb{E}\left\{\left. \Pi_B(t,T) \right| t \right\} =
\mathbb{E}\left\{\left. \sum_{i=1}^N D(t,T_i)C_i(T_i) \right| t \right\},
\end{equation}
with $T_1>t$, $T_N=T$, and $C_i$ the cashflow paid in $T_i$ that depends on the values of the risk factors (e.g. interest rates, stock prices) already known in $T_i$. We also define $\tau=\min(\tau_A,\tau_B)$, and we will use the notation $\mathbb{P}_A(t_1,t_2)=\mathbb{E}\left\{\mathbb{I}_A(t_1,t_2)\right\}$, and analogous for $B$, for the unconditioned probabilities. We will omit conditioning on $t_0$ henceforth.

The last two terms in eq.~(\ref{fairvalue}) define the Bilateral Credit Valuation Adjustment (BCVA) and Debt Valuation Adjustment (BDVA) respectively as seen from $B$,
\begin{equation}\label{BCVA}
BCVA_B(t,T) = \mathbb{E}\left\{\left. L_A\mathbb{I}_A(t,T) D(t,\tau_A) \left[ V_B^0(\tau_A) \right]^+ \right| t \right\},
\end{equation}
\begin{equation}\label{BDVA}
BDVA_B(t,T) = \mathbb{E}\left\{\left. L_B\mathbb{I}_B(t,T) D(t,\tau_B) \left[ -V_B^0(\tau_B) \right]^+ \right| t \right\},
\end{equation}
so that we can write the well known formula
\begin{equation}\label{fairvalue1}
V^{AB}_B(t_0) = V^0_B(t_0) -BCVA_B(t_0,T)+BDVA_B(t_0,T).
\end{equation}

\section{Main Results}
\subsection{Unilateral Break Clause}
Let us assume that party $B$ has a BC at time $\hat{t}<T$, i.e.~she has the right to terminate the derivative at time $\hat{t}$ by liquidating its default-free fair value $V_B^0(\hat{t})$.
It is easy to show that the equivalent of eq.~(\ref{fairvalue}) becomes
\begin{eqnarray}\label{UBC0}
\hat{V}_B^{AB}(t_0) &=& V_B^0(t_0) -\nonumber \\
&& \mathbb{E}\left\{L_A \mathbb{I}_A\left(t_0,\hat{t}\right) D(t_0, \tau_A)\left[ V_B^0(\tau_A) \right]^+  \right\}+
\mathbb{E}\left\{L_B \mathbb{I}_B\left(t_0,\hat{t}\right) D(t_0, \tau_B)\left[- V_B^0(\tau_B) \right]^+ \right\} - \\
&& \mathbb{E}\left\{\mathbb{I}_{V_B^0(\hat{t}) \leq V_B^{AB}(\hat{t})} \left(
L_A\mathbb{I}_A(\hat{t},T) D(t_0,\tau_A) \left[ V_B^0(\tau_A) \right]^+
-L_B\mathbb{I}_B(\hat{t},T) D(t_0,\tau_B) \left[ -V_B^0(\tau_B) \right]^+
\right)\right\},\nonumber
\end{eqnarray}
the exercise condition of the BC being
\begin{equation}\label{exC}
V_B^0(\hat{t}) \ge V_B^{AB}(\hat{t}).
\end{equation}
By conditioning on $\hat{t}$ the inner part of the last term in (\ref{UBC0}), we obtain our main result:
\begin{eqnarray}\label{UBC}\nonumber
\hat{V}_B^{AB}(t_0) &=& V_B^0(t_0)-BCVA_B(t_0,\hat{t})+BDVA_B(t_0,\hat{t})+	\\
&&\mathbb{E}\left\{ \mathbb{I}_{\tau>\hat{t}} D(t_0,\hat{t})
\left[ BDVA_B(\hat{t},T)-BCVA_B(\hat{t},T) \right]^+\right\}.
\end{eqnarray}
The last term is equivalent to
\begin{equation}\nonumber
\mathbb{E}\left\{ \mathbb{I}_{\tau>\hat{t}} D(t_0,\hat{t})
\left[ V_B^{AB}(\hat{t})-V_B^0(\hat{t}) \right]^+ 
\right\},
\end{equation}
where one easily recognizes the payoff linked to the continuation value of the option to liquidate the derivative in $\hat{t}$.

Equation (\ref{UBC}) has essentially the following meaning: the party that is long the BC continues the transaction at $\hat{t}$ only if its expected gain (BDVA) exceeds its expected loss (BCVA) or, equivalently, if the value of the transaction (including the bilateral valuation adjustment) is worth more than the default risk-free value. By reducing the impact of the bilateral valuation adjustment, a BC brings the value of the transaction near its default risk-free value. In particular, it reduces the expected loss of the transaction, due to a number of scenarios in which the BC is exercised and any loss due to the default of the counterparty may arise just until $\hat{t}$ instead of the maturity date. Such shortening of the credit exposure's horizon is the main reason for which also the counterparty risk in terms of CVA variability is mitigated, as we will show in the Section \ref{CVARisk} below.

\subsection{Multiple Unilateral Break Clause}
Define the last three terms of eq.~(\ref{UBC}) as $UBC(t_0, \hat{t})$,
so that we can write $\hat{V}_B^{AB}(t_0) = V_B^0(t_0)+UBC(t_0, \hat{t})$. If we add an additional BC in $\hat{t}_2> \hat{t}$ we can repeat the same arguments of the previous section to see that eq.~(\ref{UBC}) becomes
\begin{equation}\label{MBC}
\hat{V}_{B,multiple}^{AB}(t_0) = V_B^0(t_0)-
BCVA_B(t_0,\hat{t})+BDVA_B(t_0,\hat{t})+
\mathbb{E}\left\{ \mathbb{I}_{\tau>\hat{t}} D(t_0,\hat{t})
\left[ UBC(\hat{t}, \hat{t}_2) \right]^+\right\},
\end{equation}
and analogous results can be obtained for an arbitrary number of BCs.

\subsection{Mutual Break Clause}
In case of mutual BC (i.e.~both parties have the right to liquidate the position in $\hat{t}$), it is easy to see that, if both parties apply the same mark-to-market policy, the BC is exercised with certainty. In fact, the exercise condition for party $A$ is
\begin{equation}
V_A^0(\hat{t}) \ge V_A^{AB}(\hat{t}),
\end{equation}
and since $V_A^0 = -V_B^0$ and (see again \cite{brigo2}) $V_A^{AB} = -V_B^{AB}$, it becomes
\begin{equation}
V_B^0(\hat{t}) \le V_B^{AB}(\hat{t}).
\end{equation}
Hence, either $A$ or $B$ will exercise, and we can write
\begin{equation}
\hat{V}_{B,mutual}^{AB}(t_0) = V_B^0(t_0)-
BCVA_B(t_0,\hat{t})+BDVA_B(t_0,\hat{t}).
\end{equation}
Therefore, a mutual BC will be more effective than a unilateral BC in mitigating the counterparty risk of a transaction, since the former, if rationally exercised, cancels any loss arising from the default of the counterparty when it takes place after the BC date.

\section{Applications} 
As in \cite{brigo2}, we assume that the default times have a bivariate exponential distribution obtained combining exponential marginal distributions with constant default intensity $\mathbb{P}\left(\tau_X>t \right) = e^{-\lambda_X t}$ with a Gumbel copula, disallowing simultaneous defaults. The resulting survivalship probability reads
\begin{equation}
\mathbb{P}\left(\tau_A>t_A, \tau_B>t_B\right) = e^{- \left[ (\lambda_A t_A)^\theta +(\lambda_B t_B)^\theta \right]^{1/\theta}},
\end{equation}
with $\theta\in [1, +\infty )$.
Observe that Kendall's Tau $\tau_K =1-1/\theta$,
so that the independent case corresponds to $\theta=1$, and the comonotonic to $\theta=\infty$.
This particular choice of copula is common in the recent literature, and has the advantage of being analytically tractable. However, the results in terms of relevance of the unilateral BC in the pricing of derivatives have been found to be qualitatively identical to those obtained with a Gaussian copula.

Assuming independence between default events and the market risk factors, we rewrite (\ref{BCVA}) and (\ref{BDVA}) as
\begin{equation}\label{calls}
BCVA_B(t_0,T) =L_A \int_{t_0}^T d\tau_A \int_{\tau_A}^{+\infty}d\tau_B\, p(\tau_A, \tau_B)
\mathbb{E}\left\{ D(t_0,\tau_A) \left[ V_B^0(\tau_A) \right]^+  \right\},
\end{equation}
\begin{equation}\label{puts}
BDVA_B(t_0,T) =L_B \int_{t_0}^T d\tau_B \int_{\tau_B}^{+\infty}d\tau_A\, p(\tau_A, \tau_B)
\mathbb{E}\left\{  D(t_0,\tau_B) \left[ -V_B^0(\tau_B) \right]^+  \right\}.
\end{equation}
where $p(\tau_A, \tau_B)$ is the probability density of the bivariate default process.

By looking at eqs (\ref{UBC}) and (\ref{MBC}) one can see that the evaluation of the BC is strictly related to the pricing of bermudan options. Among the various techniques that can be used to solve this kind of options we will first use the Geske-Johnson technique, that is well suited only for a small number of exercise dates, and with some additional approximations leads to analytical results in the case of an equity forward. 
To generalize our results, we will then evaluate the BC on a interest rate swap (IRS) using a short rate model on a trinomial tree, a powerful technique that is widely used in the industry.

\subsection{Equity Forward with Simplified Settings}
We apply the pricing formula (\ref{UBC}) to the case of an equity forward on a non dividend paying stock $S_t$ following Black and Scholes dynamics, with strike $K$, maturity $T$ and unilateral BC at a single date $\hat{t}<T$.

To simplify the subsequent formulas let us assume that default can happen up to $\hat{t}^-$ or $T^-$, and in case of default the parties exchange the close-out amount as evaluated at $\hat{t}$ or $T$ respectively rather than at the time of default. For a payer without BC, the pricing formula (\ref{fairvalue1}) together with eqs (\ref{calls}) and (\ref{puts}) becomes
\begin{eqnarray}\label{noBC}
V_{B,payer}^{AB}(t_0)&=&S_0-KP(t_0,T)-\nonumber\\
&& L_A\mathbb{P}_A(t_0,\hat{t})\, Call(t_0;KP(\hat{t},T),\hat{t})+
L_B\mathbb{P}_B(t_0,\hat{t})\,Put(t_0;KP(\hat{t},T),\hat{t})-\nonumber\\
&& L_A\mathbb{P}_A(\hat{t},T)\,Call(t_0;K,T)+
L_B\mathbb{P}_B(\hat{t},T)\,Put(t_0;K,T),
\end{eqnarray}
where $Call(t;K,\tau)$ ($Put(t;K,\tau)$) stands for the value in $t$ of a European call (put) option with maturity $\tau$ on the contingent claim $D(\tau,T)(S_T-K)$, and $P(\hat{t},T) = \mathbb{E}(D(\hat{t},T)|\hat{t})$.
We point out that the inclusion of the terms BCVA and BDVA changes the par strike, except when the two counterparies are identical ($L_A=L_B=L$ and $\lambda_A=\lambda_B=\lambda$).

Now, introducing the BC in favor of $B$ at $\hat{t}$, we have to calculate the last expectation in (\ref{UBC0}) or (\ref{UBC}), which under our hypotheses becomes
\begin{equation}\nonumber
\mathbb{E}\left\{ D(t_0,\hat{t})\left[
L_B \mathbb{P}_B(\hat{t},T) Put(\hat{t};K,T)- L_A \mathbb{P}_A(\hat{t},T)Call(\hat{t};K,T)
\right]^+\right\},
\end{equation}
which is positive when the continuation condition
\begin{equation}
L_B \mathbb{P}_B(\hat{t},T)\,Put(\hat{t};K,T) \geq L_A \mathbb{P}_A(\hat{t},T)\,Call(\hat{t};K,T)
\end{equation}
is satisfied.
In order to apply the Geske-Johnson technique \cite{brockhaus}, we determine the boundary of such region, i.e.~we look for a $S_{\hat{t}}=U$ such that
\begin{equation}\label{condition}
L_B \mathbb{P}_B(\hat{t},T)\,Put(\hat{t};K,T)=L_A \mathbb{P}_A(\hat{t},T)\,Call(\hat{t};K,T);
\end{equation}
this allows us to write the call and put payoffs in terms of indicator functions, arriving at the final formula
\begin{eqnarray}\label{result}\nonumber
\hat{V}_{B,payer}^{AB}(t_0)&=&S_0-KP(t_0,T) -\\\nonumber
&& L_A\mathbb{P}_A(t_0,T)\,Call(t_0;KP(\hat{t},T),\hat{t})+
L_B\mathbb{P}_B(t_0,T)\,Put(t_0;KP(\hat{t},T),\hat{t})+\\\nonumber
&& L_B\mathbb{P}_B(\hat{t},T)\,\mathbb{E} \left\{ 
D(T_0,T)(K-S_T)\mathbb{I}_{S_{\hat{t}}<U}\mathbb{I}_{S_T<K} \right\}-\\
&& L_A\mathbb{P}_A(\hat{t},T)\,\mathbb{E} \left\{ 
D(T_0,T)(S_T-K)\mathbb{I}_{S_{\hat{t}}<U}\mathbb{I}_{S_T>K} \right\}.
\end{eqnarray}
It is straightforward to price analytically these barrier options within a Black and Scholes approximation, computing a two-dimensional Gaussian integral\footnote{a multiple BC with $n$ exercise dates would require a $n+1$-dimensional Gaussian integral.}. Analogous results are obtained for a receiver equity forward.

\subsection{Numerical Results - Equity Forward}
We evaluate the impact of the BC by comparing the at-the-money strike implied by eq.~(\ref{result}) with that coming from eq.~(\ref{noBC}). All the results are obtained for $S_0=1$, $\sigma_t=0.3$, $L_A=L_B=100\%$ and zero interest rates. We will report the results also for the receiver forward, that are almost perfectly antithetic to those of the payer forward (this effect is enhanced by the choice of zero interest rate and dividend yield, and of equal loss given default). We point out that this instrument is particularly sensitive to counterparty risk, since all cash flows are concentrated at the end of its life, and is therefore an ideal case to illustrate the effects of the BC.

\begin{figure}[!h]
\centering
\includegraphics[width=0.75\textwidth]{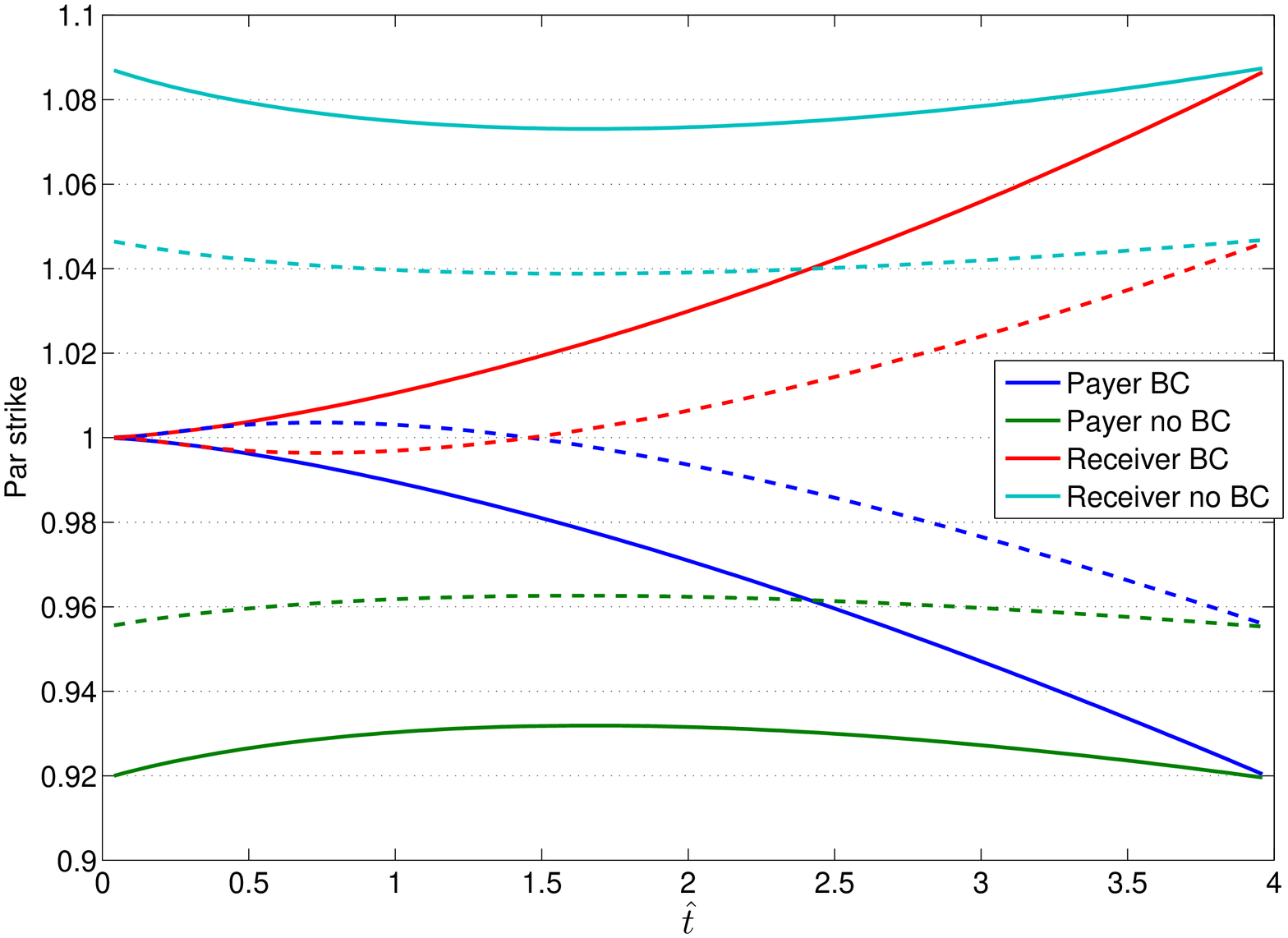}
\caption{Par strike as a function of $\hat{t}$ for a 4y equity forward with $\lambda_A=0.1$, $\lambda_B=0.05$, $\theta=4$ (full lines) and $\theta=1$ (dashed lines). As $\hat{t}$ approaches $0$ the counterparty risk is completely removed, since it is convenient to exercise the BC as soon as possible, being A the riskier counterparty.\label{fig1}}
\end{figure}

We examine first the effect of the distance between $\hat{t}$ and $T$ for $\tau_K =0$ and $\tau_K =0.75$. In figure \ref{fig1}, where $\lambda_A=0.1$, $\lambda_B=\lambda_A/2$, we see that, as $\hat{t}$ approaches $0$, the counterparty risk is completely removed (the par strike is equal to $1$, the at-the-money forward when counterparty risk is not included), since it is convenient to exercise the BC as soon as possible, being A the riskier counterparty. As $\hat{t}$ increases, the effect of the BC is reduced, and the par strike approaches the one implied by the full BCVA and BDVA contributions. When $\hat{t}$ is halfway through the life of the forward, it reduces by almost a factor of two the counterparty risk.
On the other hand, after reversing the ratio of the default intensities ($\lambda_A=0.05$, $\lambda_B=2\lambda_A$), we do not find any efficient removal of counterparty risk, an effect that is common to other examples below when the counterparty that is long the BC is more risky than the other, i.e.~when $\lambda_B>\lambda_A$.

\begin{figure}
\centering
\includegraphics[width=0.75\textwidth]{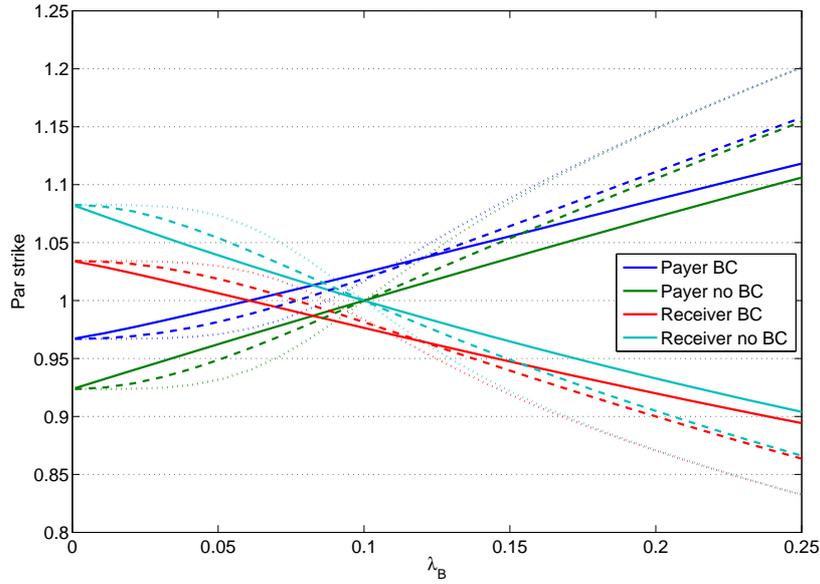}
\caption{Par strike as a function of $\lambda_B$ for an equity forward with $\lambda_A=0.1$, $T=4$, $\hat{t}=2$, $\theta=1$ (full lines), $\theta=2$ (dashed lines) and $\theta=4$ (dotted lines). 
The correction due to the BC decreases as $\lambda_B$ grows larger than $\lambda_A$ and goes to zero faster for large $\theta$. \label{figLambda}}
\end{figure}

In figure (\ref{figLambda}) we show for $\tau_K$ between 0 and 0.75 that the correction due to the BC decreases as $\lambda_B$ grows larger than $\lambda_A$ and goes to zero faster for large $\theta$. This effect can be explained by noticing that the more probable it is that $\tau_B<\tau_A$, the less useful the BC is to B.

Table (\ref{tabfig4}) shows, again with $\lambda_A=0.1$, $\lambda_B=\lambda_A/2$, how effective the BC is in reducing the credit exposure of the transaction across all values of $\theta$: the par strike of a 4 years and a 2 years equity forward with a BC after 1 year is almost identical to that of the 1 year forward. As expected, the case with $\lambda_B > \lambda_A$ (not shown) displays almost no dependence on the BC.

\begin{table}[h] 
\centering
\begin{tabular}{*{9}{|c}|}
\hline
 &\multicolumn{4}{c|}{Payer} &\multicolumn{4}{c|}{Receiver}\\
 \cline{2-9}
\rule{0pt}{3ex}$\theta$ & $T=4,\hat{t}=1$ & $T=4$& $T=2,\hat{t}=1$ & $T=2$ &
 $T=4,\hat{t}=1$ & $T=4$& $T=2,\hat{t}=1$ & $T=2$\\
\hline
1 & 0,90 & -3,23 & 0,43 & -0,81 & -0,90 & 3,37 & -0,43 & 0,82\\
2 & 0,24 & -4,42 & 0,16 & -1,09 & -0,24 & 4,70 & -0,16 & 1,12\\
3 & 0,06 & -5,33 & 0,06 & -1,31 & -0,06 & 5,75 & -0,06 & 1,36\\
4 & 0,01 & -5,91 & 0,02 & -1,45 & -0,01 & 6,42 & -0,02 & 1,51\\
5 & 0,00 & -6,23 & 0,01 & -1,53 & 0,00 & 6,80 & -0,01 & 1,59\\
\hline
\end{tabular}
\caption{Par strike difference (in percent) for various $\theta$ for $T=4$ and $T=2$ equity forwards with $\lambda_A=0.1$, $\lambda_B=0.05$, with and without a BC in $\hat{t}=1$ with respect to the 1y case with no BC. The BC reduces the 4 years and the 2 years cases to the 1y with no BC.}\label{tabfig4}
\end{table}

\subsection{Numerical Results - Plain Vanilla Swap}
To explore further the magnitude of the price change due to the introduction of a unilateral BC, we compute numerically $UBC(t_0, \hat{t})$ for a 4 year plain vanilla par interest rate swap (IRS) with semi-annual payments using a 1 factor Hull \& White model with piecewise constant volatility on a trinomial tree calibrated on market data from August 13th 2010, when the 4 years par swap rate vs Euribor 6m was $1.677\%$ and the 2y2y EUR ATM swaption volatility was $37.6\%$. The trinomial tree allows for a more accurate computation of the $BCVA(t_1,t_2)$ and $BDVA(t_1,t_2)$. In fact, it is possible to discretize eqs (\ref{calls}) and (\ref{puts}) to get
\begin{equation}
BCVA_B(t_0,T) = \sum_{k=1}^N L_A \mathbb{P}_A(T_{k-1},T_k)
\mathbb{E}\left\{ D(t_0,T_{k-1}) \left[ V_B^0(T_{k-1}) \right]^+ \right\},
\end{equation}
\begin{equation}
BDVA_B(t_0,T) = \sum_{k=1}^N L_B \mathbb{P}_B(T_{k-1},T_k)
\mathbb{E}\left\{ D(t_0,T_{k-1}) \left[ -V_B^0(T_{k-1}) \right]^+ \right\},
\end{equation}
where the $\{T_k\}_{k=1,...,N}$ are the dates in which some cash flow is exchanged, with $T_0= t_0$ and $T_N = T$.

\begin{figure}[!h]
\centering
\includegraphics[width=0.80\textwidth]{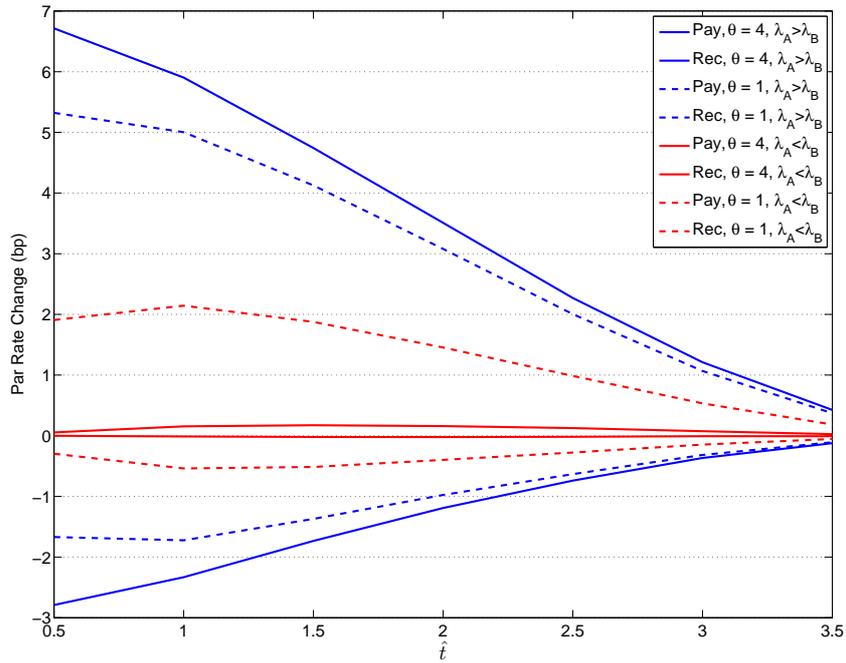}
\caption{Par rate change with respect to the no BC case as a function of $\hat{t}$ for a 4y payer (positive) and receiver (negative) IRS with $\lambda_A=0.1$, $\lambda_B=0.05$ (blue) and $\lambda_A=0.05$, $\lambda_B=0.1$ (red), $\theta=4$ (full lines) and $\theta=1$ (dashed lines). The effect of the BC is larger for $\hat{t}$ close to $0$, for $\lambda_A <\lambda_B$ and for the payer.
\label{fig1b}}
\end{figure}

\begin{figure}
\centering
\includegraphics[width=0.75\textwidth]{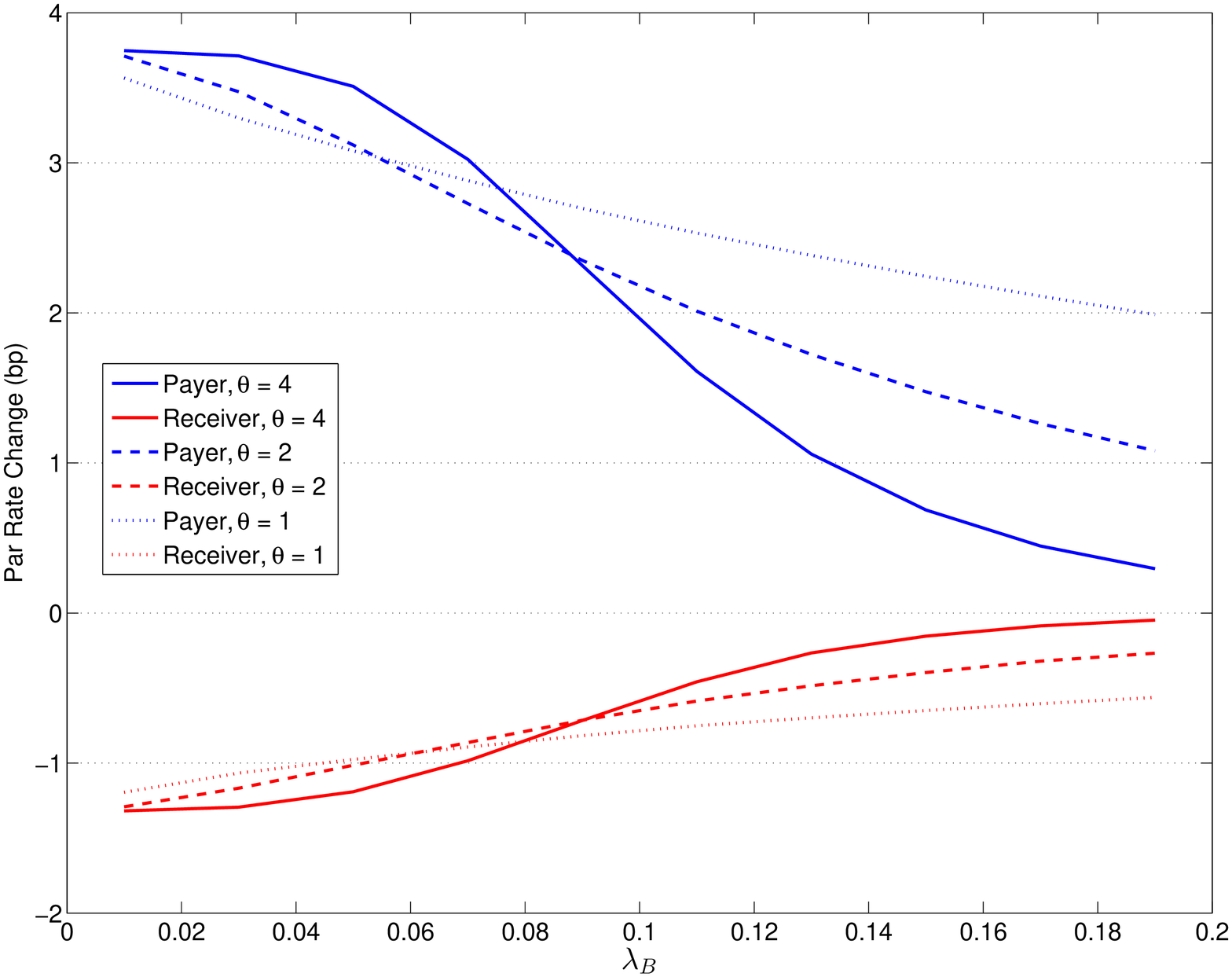}
\caption{Par rate change with respect to the no BC case as a function of $\lambda_B$ for a payer (blue) and receiver (red) IRS with $\lambda_A=0.1$, $T=4$, $\hat{t}=2$, $\theta=4$ (full lines), $\theta=2$ (dashed lines) and $\theta=1$ (dotted lines). 
The correction due to the BC decreases as $\lambda_B$ grows larger than $\lambda_A$ and goes to zero faster for large $\theta$.
\label{fig3b}}
\end{figure}

The results for both payer and receiver swap with $L_A=L_B=100\%$, expressed in terms of the change in par rate between a swap with no BC and the same swap after the introduction of one BC, are displayed in figure \ref{fig1b} and \ref{fig3b}. They are qualitatively similar to those for the equity forward, and contribute up to a few basis points to the par rate. Compared with the bid-ask spread of about 1 bp usually encountered in the OTC market for plain vanilla swaps, this amount is quite significant, even though the same spread for uncollateralized deals is usually larger. We notice that, since in this case interest rates are not zero, there is a strong asymmetry between payer and receiver swaps.

\subsection{Multiple Break Clauses}
We now investigate numerically the effect of multiple BCs in the case of the 4y IRS above. The results are collected in table \ref{tabMBC}, where in column $\hat{t}$ we show the times (in years) at which there is a BC. The main contribution comes always from the first BC, while the subsequent ones have an impact that is at least one order of magnitude smaller. This effect is enhanced by the use of constant default intensity. We argue that the introduction of a dynamics for the default probability should lead to different results, as a consequence of the possible inversion of the default order during the life of the derivative.

\begin{table}[h] 
\centering
\begin{tabular}{*{4}{|c}*{2}{|>{\centering\arraybackslash}m{1.6cm}}|*{4}{|c}*{2}{|>{\centering\arraybackslash}m{1.6cm}}|}
\hline
$\hat{t}$& $\lambda_A$& $\lambda_B$& $\theta$& Payer swap (bp) & Receiver swap (bp)&
$\hat{t}$& $\lambda_A$& $\lambda_B$& $\theta$& Payer swap (bp) & Receiver swap (bp)\\
\hline
1 2 3 & 0,1 & 0,05 & 4 & 6,0 & -2,4 & 1 2 3 & 0,05 & 0,1 & 4 & 0,2 & 0\\
1 2 & 0,1 & 0,05 & 4 & 5,9 & -2,4 & 1 2 & 0,05 & 0,1 & 4 & 0,2 & 0\\
1 & 0,1 & 0,05 & 4 & 5,9 & -2,3 & 1 & 0,05 & 0,1 & 4 & 0,2 & 0\\
2 3 & 0,1 & 0,05 & 4 & 3,5 & -1,2 & 2 3 & 0,05 & 0,1 & 4 & 0,2 & 0\\
2 & 0,1 & 0,05 & 4 & 3,5 & -1,2 & 2 & 0,05 & 0,1 & 4 & 0,2 & 0\\
3 & 0,1 & 0,05 & 4 & 1,2 & -0,4 & 3 & 0,05 & 0,1 & 4 & 0,1 & 0\\
\hline
1 2 3 & 0,1 & 0,05 & 1 & 5,3 & -2 & 1 2 3 & 0,05 & 0,1 & 1 & 2,4 & -0,7\\
1 2 & 0,1 & 0,05 & 1 & 5,3 & -1,9 & 1 2 & 0,05 & 0,1 & 1 & 2,4 & -0,7\\
1 & 0,1 & 0,05 & 1 & 5,0 & -1,7 & 1 & 0,05 & 0,1 & 1 & 2,1 & -0,5\\
2 3 & 0,1 & 0,05 & 1 & 3,1 & -1 & 2 3 & 0,05 & 0,1 & 1 & 1,5 & -0,4\\
2 & 0,1 & 0,05 & 1 & 3,1 & -1 & 2 & 0,05 & 0,1 & 1 & 1,5 & -0,4\\
3 & 0,1 & 0,05 & 1 & 1,1 & -0,3 & 3 & 0,05 & 0,1 & 1 & 0,5 & -0,1\\
\hline
\end{tabular}
\caption{Effect of multiple BCs for a 4 years swap. See text for parameters' values. The par rate change with respect to the no-BC case is driven essentially by the time to the first BC.}\label{tabMBC}
\end{table}

\subsection{CVA risk}\label{CVARisk}
Having shown the effectiveness of BCs in reducing the BCVA through a mechanism that shortens the credit exposure of the derivative, we evaluate here its impact on the CVA risk. Basel III has introduced a new capital charge to cover the CVA risk, which is measured in terms of the sensitivity of the UCVA with respect to the CDS of the counterparty \cite{basel3}. We therefore compute numerically the sensitivity of the UCVA with respect to the counterparty default intensity $\lambda_A$, both with and without a unilateral BC, as this is a good proxy of the sensitivity defined by the Basel Committee, using the well known market formula $\lambda_A LGD \approx CDS$. The results for the 4y IRS described previously, collected in table \ref{TCVARisk}, show that the BC significantly reduces also this risk measure.

\begin{table}[h]
\centering
\begin{tabular}{*{3}{|c}*{4}{|>{\centering\arraybackslash}m{1.8cm}}|}
\hline
\rule{0pt}{2.5ex} 
& & &\multicolumn{2}{c|}{Payer UCVA sensitivity} &\multicolumn{2}{c|}{Receiver UCVA  sensitivity}\\
\cline{4-7}
\rule{0pt}{2.5ex} $\lambda_A$& $\lambda_B$&	$\theta$& with BC& no BC & with BC& no BC \\
\hline
0,05 & 0,1 & 4 & 1,634 & 2,812 & 1,094 & 1,321\\
0,05 & 0,1 & 1 & 1,459 & 2,812 & 0,892 & 1,321\\
0,1 & 0,05 & 4 & 1,209 & 2,244 & 0,728 & 1,089\\
0,1 & 0,05 & 1 & 1,234 & 2,244 & 0,748 & 1,089\\
0,1 & 0,1 & 4 & 1,260 & 2,244 & 0,753 & 1,089\\
0,1 & 0,1 & 1 & 1,260 & 2,244 & 0,753 & 1,089\\
\hline
\end{tabular}
\caption{Comparison of sensitivities of UCVA with respect to $\lambda_A$ (in bps) with and without BC.}\label{TCVARisk}
\end{table}

\section{Conclusions}
We have shown that the financial effect of the BC (both unilateral and bilateral) is quite significant for various kinds of plain vanilla derivatives and a large range of parameters. The impact is a consequence the use of the default risk-free close-out amount, which is legitimate according to the ISDA close out protocol. In particular, the BC mitigates the CVA risk of a transaction and the related Basel III capital charges.

This result must however be contrasted with the historically low propensity of counterparties to exercise the BC in order not to compromise their relationship with clients. 

Our approach can be generalized to more complex BCs, for instance conditioned to rating downgrade, that we leave for future work. Another development will be to investigate the effect of stochastic default rates that will add volatility to the credit adjustment and also alter the order of default of the counterparties by weakening the dependence of the result on the ratio between $\lambda_B$ and $\lambda_A$ that is kept constant in our examples.

\begin{thebibliography}{99}
\bibitem{brigo3} \textsc{D. Brigo, M. Morini}, {\it Close-out convention tensions}, {\it Risk} December 2011

\bibitem{brigo2} \textsc{D. Brigo, C. Buescu, M. Morini}, {\it Impact on the first to default time on a Bilateral CVA}, June 20, 2011, available at \textsc{http://www.defaultrisk.com}

\bibitem{mercurio} \textsc{F. Mercurio, R. Caccia and M. Cutuli}, {\it Downgrade termination costs}, {\it Risk} April 2012

\bibitem{riskmagazine1} \textsc{L. Carver}, {\it ISDA to review close-out value definitions}, {\it Risk online} September 2011

\bibitem{riskmagazine2} \textsc{L. Carver}, {\it CVA's cousin: Dealers try to value early termination clauses}, {\it Risk} September 2011

\bibitem{riskmagazine3} \textsc{M. Cameron}, {\it Italy could face more swap terminations}, {\it Risk online} March 2012

\bibitem{riskmagazine4} \textsc{M. Cameron}, {\it Dealers call for revised language on break clause close-outs}, {\it Risk online} April 2012

\bibitem{riskmagazine5} \textsc{M. Cameron}, {\it Breaking with tradition}, {\it Risk} March 2012

\bibitem{brockhaus} \textsc{O. Brockhaus, A. Ferraris, C. Gallus, D. Long, R. Martin, M. Overhaus}, {\it Modelling and hedging equity derivatives}, 1999, \textsc{Risk Books}

\bibitem{isda} \textsc{International Swap and Derivatives Association}, {\it ISDA Close-out Amount Protocol}, February 27, 2009, available at \textsc{http://www.isda.org/isdacloseoutamtprot/docs/isdacloseoutprot-text.pdf}

\bibitem{basel3}\textsc{Basel Committee on Banking Supervision}, {\it Basel III: A global regulatory framework for more resilient banks and banking systems}, December 2010 (rev. June 2011), available at \textsc{http://www.bis.org/publ/bcbs189.pdf}

\end{thebibliography}
\end{document}